\documentclass[preprint,prd,showpacs,showkeys]{revtex4-1}
\usepackage{amsmath}
\usepackage{amssymb}
\usepackage{amsfonts}
\usepackage{graphicx}
\usepackage{appendix}

\begin{document}

\title{Muon Anomalous Magnetic Moment and Gauge Symmetry in the Standard
Model}
\author{Er-Cheng Tsai}
\email{ectsai@ntu.edu.tw}
\affiliation{Physics Department, National Taiwan University, Taipei, Taiwan}

\begin{abstract}
No gauge invariant regularization is available for the perturbative
calculation of the standard model. One has to add finite counter terms to
restore gauge symmetry for the renormalized amplitudes. The muon anomalous
magnetic moment can be accurately measured but the experimental result does
not entirely agree with the theoretical calculation from the standard model.
This paper is to compute the contributions to the muon gyromagnetic ratio $%
g_{\mu}$ due to the finite counter terms. The result obtained is found to be
far from sufficient to explain the discrepancy between theory and experiment.
\end{abstract}

\maketitle
\affiliation{Physics Department, National Taiwan University, Taipei, Taiwan}

\section{Introduction}

The magnetic moment of a muon is $\vec{M}=g_{\mu}\frac{e}{2m_{\mu}}\vec{S}$
where the gyromagnetic ratio $g_{\mu}$ is equal to $2$ if quantum loop
corrections are ignored. $g_{\mu}$ can be measured quite precisely and its
derivation from the classical value is found to be $\frac{g_{\mu}-2}%
{2}=11659209.1\times10^{-10}$ \cite{MUG2, MUgexp}. Within the framework of the
standard model, $\frac{g_{\mu}-2}{2}$ can be calculated and its theoretical
value compared to the experimental result therefore offers a precise test of
the standard model at quantum loop level.

The electromagnetic form factors can be written as:%
\[%
\raisebox{-0.4289in}{\includegraphics[
height=1.0018in,
width=1.6143in
]%
{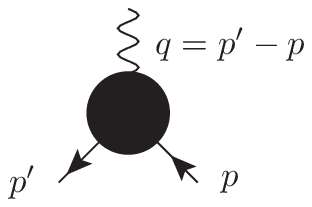}%
}
=\bar{u}\left(  p^{\prime}\right)  \left(  \gamma^{\mu}+a_{\mu}\frac
{1}{2m_{\mu}}\left[  \not q,\frac{1}{2}\gamma^{\mu}\right]  \right)  u\left(
p\right)  +O\left(  q^{2}\right)
\]
with $p^{2}=p^{\prime2}=m_{\mu}^{2}$ and $a_{\mu}=\frac{g_{\mu}-2}{2}$
\cite{FFR}. In the standard model higher order corrections of $a_{\mu}$ are
classified into QED, electroweak (EW) and hadronic classes:%
\[
a_{\mu}=a_{\mu}^{QED}+a_{\mu}^{EW}+a_{\mu}^{Had}%
\]
The QED part is known to 4-loops and leading terms in 5-loops \cite{QED5loop}.%
\begin{align}
a_{\mu}^{QED} &  =\left(  \frac{\alpha}{\pi}\right)  0.5+\left(  \frac{\alpha
}{\pi}\right)  ^{2}0.765857410(27)+\left(  \frac{\alpha}{\pi}\right)
^{3}24.05050964(87)\nonumber\\
&  +\left(  \frac{\alpha}{\pi}\right)  ^{4}130.8055(80)+\left(  \frac{\alpha
}{\pi}\right)  ^{5}663(20)\nonumber\\
&  =116584718.09(0.15)\times10^{-11}\label{aQED}%
\end{align}
The electroweak part $a_{\mu}^{EW}$ is the loop contribution due to heavy
$W^{\pm}$, $Z$ or Higgs particle and is suppressed by at least a factor of%
\[
\frac{\alpha}{\pi}\frac{m_{\mu}^{2}}{m_{W}^{2}}\simeq4\times10^{-9},
\]
which enables us to neglect the 3-loop terms. The 1-loop \cite{aem1}%
\[
a_{\mu}^{EW}\left[  \text{1-loop}\right]  =194.8\times10^{-11}%
\]
and leading term in 2-loop \cite{aem2}%
\[
a_{\mu}^{EW}\left[  \text{2-loop}\right]  =-40.7\times10^{-11}%
\]
add up to give the total%
\begin{equation}
a_{\mu}^{EW}=154\times10^{-11}.\label{aEW}%
\end{equation}
The hadronic part is evaluated via dispersion relation approach, the available
$\sigma\left(  e^{+}e^{-}\rightarrow hadrons\right)  $ data give rise
\cite{HADEPM} to a leading-order hadronic vacuum polarization contribution of
\cite{HADLO}%
\begin{equation}
a_{\mu}^{Had}\left[  LO\right]  =6923\left(  42\right)  \left(  3\right)
\times10^{-11}.\label{aHadLO}%
\end{equation}
Higher order hadronic contribution is found to be \cite{HADNLO}%
\begin{equation}
a_{\mu}^{Had}\left[  NLO\right]  =7\left(  26\right)  \times10^{-11}%
\label{aHadNLO}%
\end{equation}
Adding $\left(  \ref{aQED}\right)  $, $\left(  \ref{aEW}\right)  $, $\left(
\ref{aHadLO}\right)  $\ and $\left(  \ref{aHadNLO}\right)  $ gives the
standard model prediction based on $e^{+}e^{-}$ data.%
\[
a_{\mu}^{SM}=116591803\left(  1\right)  \left(  42\right)  \times10^{-11}%
\]
The difference between experiment and theory is%
\begin{equation}
\Delta a_{\mu}=a_{\mu}^{\exp}-a_{\mu}^{SM}=281\left(  63\right)  \left(
49\right)  \times10^{-11}\label{gdiff}%
\end{equation}
New physics effects beyond standard model have been pondered over to explain
this discrepancy.

The dimensional regularization scheme proposed by 't Hooft and Veltman
\cite{HV} is a very convenient scheme for regularizing gauge theory without
$\gamma_{5}$, such as QED. For chiral gauge theories involving $\gamma_{5}$,
no gauge invariant regularization is available but the dimensional
regularization can still be used in a rigorous manner by maintaining
\begin{equation}
\gamma_{5}=i\gamma^{0}\gamma^{1}\gamma^{2}\gamma^{3} \label{g5def}%
\end{equation}
even when the space-time dimension $n$ departs from $4$ \cite{BM}. Such
$\gamma_{5}$ anticommutes with $\gamma^{\mu}$ for $\mu$\ in the first 4
dimensions but commutes with $\gamma^{\mu}$ when the index $\mu$ goes beyond
the first 4 dimensions. The continuation to $n\neq4$ for the Lagrangian of a
theory with a gauge invariant 4 dimensional Lagrangian depends on how we
express and continue the terms involving $\gamma_{5}$ in the Lagrangian. The
breakdown of gauge symmetry in this scheme can be remedied by introducing
additional finite gauge variant local counter terms to restore the gauge
symmetry \cite{FCTMS}. One ingredient that was missing in the previous
evaluation of the electroweak part $a_{\mu}^{EW}$ is the contribution due to
these finite counter terms that must be invoked to restore gauge symmetry. In
this paper, we will calculate the correction to the muon gyromagnetic ratio
due to the lowest order finite counter terms. It turns out that the result
obtained is not significant enough to account for the difference $\left(
\ref{gdiff}\right)  $.

\section{Standard Model}

The gauge group for the standard model is $SU\left(  3\right)  \times
SU\left(  2\right)  \times U\left(  1\right)  $ with three kinds of vector
gauge bosons, $G^{\mu,a},a=1,2,..8$ for $SU\left(  3\right)  $, $W^{\mu,a},$
$a=1,2,3$ for $SU\left(  2\right)  $, and $B^{\mu}$ for $U\left(  1\right)  $.
Let $S^{a},a=1,2,..8$ and $T^{a},a=1,2,3$ be the traceless, Hermitian
generators for $SU\left(  3\right)  $\ and $SU\left(  2\right)  $ in the
adjoint representation. They are normalized as%
\begin{equation}
Tr\left(  S^{a}S^{b}\right)  =\frac{1}{2}\delta^{ab},Tr\left(  T^{a}%
T^{b}\right)  =\frac{1}{2}\delta^{ab} \label{gnmc}%
\end{equation}
with the commutators%
\begin{equation}
\left[  S^{a},S^{b}\right]  =if^{abc}S^{c},\left[  T^{a},T^{b}\right]
=i\epsilon^{abc}T^{c}\nonumber
\end{equation}
We choose $T^{a}=\frac{\sigma_{a}}{2}$ as the $SU\left(  2\right)  $ generator
with $\sigma_{a}$ being the Pauli matrix. Define the matrix fields%
\[
G^{\mu}=\sum_{a=1}^{8}G_{a}^{\mu}S^{a},W^{\mu}=\sum_{a=1}^{8}W_{a}^{\mu}T^{a}%
\]
and the covariant derivatives%
\[
D_{S}^{\mu}=\partial^{\mu}+ig_{S}G^{\mu},D_{W}^{\mu}=\partial^{\mu}%
+ig_{W}W^{\mu},D_{B}^{\mu}=\partial^{\mu}+i\frac{g_{B}}{2}B^{\mu},
\]
for $SU\left(  3\right)  ,SU\left(  2\right)  $ and $U\left(  1\right)  $ with
coupling constants $g_{S},g_{W}$ and $g_{B}$ respectively. Let%
\[
G^{\mu\nu}=\frac{1}{ig_{S}}\left[  D_{S}^{\mu},D_{S}^{\nu}\right]
=\partial^{\mu}G^{\nu}-\partial^{\nu}G^{\mu}+ig_{S}\left[  G^{\mu},G^{\nu
}\right]  ,
\]%
\[
W^{\mu\nu}=\frac{1}{ig_{W}}\left[  D_{W}^{\mu},D_{W}^{\nu}\right]
=\partial^{\mu}W^{\nu}-\partial^{\nu}W^{\mu}+ig_{W}\left[  W^{\mu},W^{\nu
}\right]  ,
\]
and%
\[
B^{\mu\nu}=\partial^{\mu}B^{\nu}-\partial^{\nu}B^{\mu}.
\]
The Lagrangian for the standard model without matter fields is%
\begin{align}
L_{1}  &  =-{\frac{1}{2}}Tr\left(  G_{\mu\nu}G^{\mu\nu}\right)  -{\frac{1}%
{2}Tr}\left(  W_{\mu\nu}W^{\mu\nu}\right)  -{\frac{1}{4}}B_{\mu\nu}B^{\mu\nu
}\label{L1}\\
&  +\left(  D_{H}^{\mu}\phi\right)  ^{\dagger}\left(  D_{H\mu}\phi\right)
-\frac{\lambda}{8}g^{2}\left(  \phi^{\dagger}\phi-\frac{v^{2}}{2}\right)
^{2}\nonumber
\end{align}
where the Higgs $\phi$ is a two component scalar complex field coupled to $W$
and $B$ gauge bosons with%
\[
D_{H}^{\mu}\phi=\left(  \partial^{\mu}+ig_{W}W^{\mu}-i\frac{g_{B}}{2}B^{\mu
}\right)  \phi
\]
$\phi$ is assumed to have the vacuum expectation value:
\[
\left\langle \phi\right\rangle =\frac{1}{\sqrt{2}}\left[
\begin{array}
[c]{c}%
0\\
v
\end{array}
\right]
\]
Express $\phi$ in terms of four real components $H$ and $\phi_{a}$, $a=1,2,3$:%
\begin{align}
\phi &  =\frac{1}{\sqrt{2}}\left[
\begin{array}
[c]{c}%
i\phi_{1}+\phi_{2}\\
H+v-i\phi_{3}%
\end{array}
\right]  =\frac{1}{\sqrt{2}}\left[
\begin{array}
[c]{cc}%
H+v+i\phi_{3} & i\phi_{1}+\phi_{2}\\
i\phi_{1}-\phi_{2} & H+v-i\phi_{3}%
\end{array}
\right]  \left[
\begin{array}
[c]{c}%
0\\
1
\end{array}
\right] \label{defsmphi}\\
&  =\frac{1}{\sqrt{2}}\left(  H+v+i\phi_{a}\sigma^{a}\right)  \left[
\begin{array}
[c]{c}%
0\\
1
\end{array}
\right]  =\hat{\phi}\left[
\begin{array}
[c]{c}%
0\\
1
\end{array}
\right] \nonumber
\end{align}
where $\hat{\phi}$ is defined as
\[
\hat{\phi}=\frac{1}{\sqrt{2}}\left(  H+v+i\phi_{a}\sigma_{a}\right)  .
\]
Note that%
\[
\hat{\phi}\left[
\begin{array}
[c]{c}%
1\\
0
\end{array}
\right]  =\frac{1}{\sqrt{2}}\left[
\begin{array}
[c]{c}%
H+v+i\phi_{3}\\
i\phi_{1}-\phi_{2}%
\end{array}
\right]  =i\sigma_{2}\left(  \frac{1}{\sqrt{2}}\left[
\begin{array}
[c]{c}%
i\phi_{1}+\phi_{2}\\
H+v-i\phi_{3}%
\end{array}
\right]  \right)  ^{\ast}=i\sigma_{2}\hat{\phi}^{\ast}\left[
\begin{array}
[c]{c}%
0\\
1
\end{array}
\right]
\]
Under a $SU\left(  2\right)  \times U\left(  1\right)  $ transformation%
\begin{equation}
\phi=\hat{\phi}\left[
\begin{array}
[c]{c}%
0\\
1
\end{array}
\right]  \rightarrow e^{-ig_{W}\theta_{a}T_{a}}e^{i\frac{g_{B}}{2}\chi}%
\hat{\phi}\left[
\begin{array}
[c]{c}%
0\\
1
\end{array}
\right]  \label{trh2}%
\end{equation}
and, since $\left(  i\sigma_{2}\right)  \vec{\sigma}^{\ast}=-\vec{\sigma
}\left(  i\sigma_{2}\right)  ,$%
\begin{equation}
\hat{\phi}\left[
\begin{array}
[c]{c}%
1\\
0
\end{array}
\right]  \rightarrow i\sigma_{2}e^{ig_{W}\theta_{a}T_{a}^{\ast}}%
e^{-i\frac{g_{B}}{2}\chi}\hat{\phi}^{\ast}\left[
\begin{array}
[c]{c}%
0\\
1
\end{array}
\right]  =e^{-ig_{W}\theta_{a}T_{a}}e^{-i\frac{g_{B}}{2}\chi}\hat{\phi}\left[
\begin{array}
[c]{c}%
1\\
0
\end{array}
\right]  \label{trh1}%
\end{equation}
Replacing $\phi$ by $\frac{1}{\sqrt{2}}\left[
\begin{array}
[c]{c}%
0\\
v
\end{array}
\right]  $ in $\left(  D_{H}^{\mu}\phi\right)  ^{\dagger}\left(  D_{H\mu}%
\phi\right)  $, we obtain the following quadratic mass term for the vector
bosons.
\begin{align*}
&  \frac{v^{2}}{2}\left[
\begin{array}
[c]{cc}%
0 & 1
\end{array}
\right]  \left(  g_{W}W^{\mu}-\frac{g_{B}}{2}B^{\mu}\right)  ^{2}\left[
\begin{array}
[c]{c}%
0\\
1
\end{array}
\right] \\
&  =\frac{v^{2}}{2}\left(  \left(  \frac{g_{W}}{2}\right)  ^{2}\left(  \left(
W_{1}^{\mu}\right)  ^{2}+\left(  W_{2}^{\mu}\right)  ^{2}\right)  +\left(
\frac{g_{W}}{2}W_{3}^{\mu}+\frac{g_{B}}{2}B^{\mu}\right)  ^{2}\right)
\end{align*}
Define%
\begin{equation}
\left[
\begin{array}
[c]{c}%
A^{\mu}\\
Z^{\mu}%
\end{array}
\right]  =\frac{1}{\sqrt{g_{W}^{2}+g_{B}^{2}}}\left[
\begin{array}
[c]{cc}%
g_{W} & -g_{B}\\
g_{B} & g_{W}%
\end{array}
\right]  \left[
\begin{array}
[c]{c}%
B^{\mu}\\
W^{\mu}%
\end{array}
\right]  . \label{AZdef}%
\end{equation}
The vector field $A^{\mu}$ is massless and is identified as the photon field.

The Lagrangian $\left(  \ref{L1}\right)  $ is invariant under the following
BRST \cite{BRST} variations with Grassmann ghost fields $c_{S}=\sum_{a=1}%
^{8}c_{S}^{a}S^{a}$, $c_{W}=\sum_{a=1}^{3}c_{W}^{a}T^{a}$, $c_{B}$ as the
parameters for the $SU\left(  3\right)  $, $SU\left(  2\right)  $, $U\left(
1\right)  $ groups.%
\begin{align}
\delta G^{\mu}  &  =\left[  D_{S}^{\mu},c_{S}\right]  ,\delta W^{\mu}=\left[
D_{W}^{\mu},c_{W}\right]  ,\delta B_{\mu}=\partial_{\mu}c_{B},\label{brsm1}\\
\delta\phi &  =-i\left(  g_{W}c_{W}-\frac{g_{B}}{2}c_{B}\right)  \phi\nonumber
\end{align}
The gauge fixing and corresponding ghost terms \cite{CTBRS} in the pure alpha
gauge are
\begin{align}
L_{gf}  &  =-\frac{1}{\alpha_{S}}Tr\left(  \partial_{\mu}G^{\mu}\right)
^{2}-\frac{1}{\alpha_{W}}Tr\left(  \partial_{\mu}W^{\mu}\right)  -\frac
{1}{2\alpha_{B}}\left(  \partial_{\mu}B^{\mu}\right)  ^{2}\label{smgf}\\
&  +2Tr\left(  i\bar{c}_{S}\delta\left(  \partial_{\mu}G^{\mu}\right)
\right)  +2Tr\left(  i\bar{c}_{W}\delta\left(  \partial_{\mu}W^{\mu}\right)
\right)  +i\bar{c}_{B}\delta\left(  \partial_{\mu}B^{\mu}\right) \nonumber
\end{align}
where $\bar{c}_{S}$, $\bar{c}_{W}$, $\bar{c}_{B}$ are the anti-ghosts
corresponding to $c_{S}$, $c_{W}$, $c_{B}$ and the BRST variations for ghost
and anti-ghost fields are%
\begin{align*}
\delta c_{S}^{a}  &  =\frac{g_{S}}{2}f^{abc}c_{S}^{b}c_{S}^{c},\delta
c_{W}^{a}=\frac{g_{W}}{2}\epsilon^{abc}c_{W}^{b}c_{W}^{c},\delta c_{B}=0,\\
\delta\bar{c}_{S}  &  =-\frac{i}{\alpha_{S}}\partial_{\mu}G^{\mu},\delta
\bar{c}_{W}=-\frac{i}{\alpha_{W}}\partial_{\mu}W^{\mu},\delta\bar{c}%
_{B}=-\frac{i}{\alpha_{B}}\partial_{\mu}B^{\mu}.
\end{align*}
There are three generations of fermion matter fields consisting of quarks%
\[
\left(
\begin{array}
[c]{c}%
u\\
d
\end{array}
\right)  ,\left(
\begin{array}
[c]{c}%
c\\
s
\end{array}
\right)  ,\left(
\begin{array}
[c]{c}%
t\\
b
\end{array}
\right)
\]
and leptons
\begin{equation}
\left(
\begin{array}
[c]{c}%
\nu_{e}\\
e
\end{array}
\right)  ,\left(
\begin{array}
[c]{c}%
\nu_{\mu}\\
\mu
\end{array}
\right)  ,\left(
\begin{array}
[c]{c}%
\nu_{\tau}\\
\tau
\end{array}
\right)  \label{lpfmrtx}%
\end{equation}
Note for simplicity, we have suppressed the color indices of quarks. We will
use the notation $\psi_{i}=\left(
\begin{array}
[c]{c}%
\psi_{i}^{u}\\
\psi_{i}^{d}%
\end{array}
\right)  $ indexed by $i$ to denote the above fermion fields. The $G^{\mu}$
gluons couple only to the quark fields with equal strength for left-handed and
right-handed quarks. $W$ and $B$ gauge bosons couple to both left-handed
quarks and left-handed leptons. The chiral projection operators $L$ and $R$
are defined as%
\[
L=\frac{1}{2}\left(  1-\gamma_{5}\right)  ,R=\frac{1}{2}\left(  1+\gamma
_{5}\right)  .
\]
The right-handed fermion $\psi_{R}=R\psi$ is a $SU\left(  2\right)  $ singlet
and thus is not coupled to $W$. The covariant derivative for a left-handed
quark $\psi_{L}=L\psi$ is%
\begin{equation}
D_{q,L}^{\mu}L\psi_{i}=\left(  \partial^{\mu}+ig_{S}G^{\mu}+ig_{W}W^{\mu
}-iY_{i}\frac{g_{B}}{2}B^{\mu}\right)  L\psi_{i} \label{cdlq}%
\end{equation}
and that for a left-handed lepton is%
\begin{equation}
D_{l,L}^{\mu}L\psi_{i}=\left(  \partial^{\mu}+ig_{W}W^{\mu}-iY_{i}\frac{g_{B}%
}{2}B^{\mu}\right)  L\psi_{i} \label{cdll}%
\end{equation}
where $Y_{i}$ is the weak hypercharge. Using the inverse of $\left(
\ref{AZdef}\right)  $ to expresses $\left(  B,W\right)  $ in terms of $\left(
A,Z\right)  $, we get%
\begin{align*}
g_{W}W_{3}^{\mu}T^{3}-\frac{g_{B}}{2}Y_{i}B^{\mu}  &  =\left(  \frac{g_{W}}%
{2}W_{3}^{\mu}+\frac{g_{B}}{2}B^{\mu}\right)  \sigma_{3}-\left(  Y_{i}%
+\sigma_{3}\right)  \frac{g_{B}}{2}B^{\mu}\\
&  =-\frac{\left(  Y_{i}+\sigma_{3}\right)  }{2}\frac{g_{B}g_{W}}{\sqrt
{g_{W}^{2}+g_{B}^{2}}}A^{\mu}\\
&  +\frac{g_{W}^{2}\sigma_{3}-Y_{i}g_{B}^{2}}{2\sqrt{g_{W}^{2}+g_{B}^{2}}%
}Z^{\mu},
\end{align*}
The electric charge for the left-handed fermion is proportional to$\frac
{\left(  Y_{i}+\sigma_{3}\right)  }{2}g_{B}$. The weak hypercharge for the
right-handed fermions must also be $\frac{\left(  Y_{i}+\sigma_{3}\right)
}{2}g_{B}$ so that the electric charges for the left-handed and right-handed
fermions may be the same. The covariant derivative for a right-handed quark is%
\begin{equation}
D_{q,R}^{\mu}R\psi_{i}=\left(  \partial^{\mu}+ig_{S}G^{\mu}-i\frac{\left(
Y_{i}+\sigma_{3}\right)  }{2}g_{B}B^{\mu}\right)  R\psi_{i} \label{cdrq}%
\end{equation}
and that for a right-handed lepton is%
\begin{equation}
D_{l,R}^{\mu}R\psi_{i}=\left(  \partial^{\mu}-i\frac{\left(  Y_{i}+\sigma
_{3}\right)  }{2}g_{B}B^{\mu}\right)  R\psi_{i}. \label{cdrl}%
\end{equation}
It is known that $Y_{i}=-1$ for all leptons and $Y_{i}=\frac{1}{3}$ for all quarks.

The transformations $\left(  \ref{trh2}\right)  $ and $\left(  \ref{trh1}%
\right)  $ for $\hat{\phi}$ can be utilized to show that the following four
types of Yukawa terms%
\[
\bar{\psi}_{i}^{d}\left(  \hat{\phi}\left[
\begin{array}
[c]{c}%
0\\
1
\end{array}
\right]  \right)  ^{\dagger}L\psi_{j},\bar{\psi}_{i}\hat{\phi}\left[
\begin{array}
[c]{c}%
0\\
1
\end{array}
\right]  R\psi_{j}^{d},\bar{\psi}_{i}^{u}\left(  \hat{\phi}\left[
\begin{array}
[c]{c}%
1\\
0
\end{array}
\right]  \right)  ^{\dagger}L\psi_{j},\bar{\psi}_{i}\hat{\phi}\left[
\begin{array}
[c]{c}%
1\\
0
\end{array}
\right]  R\psi_{j}^{u}%
\]
are gauge invariant provided the $\psi_{i}$ and $\psi_{j}$ fields in the above
have the same weak hypercharge. The Yukawa interaction for quarks can be
written as%
\begin{equation}
L_{YQ}=-\sum_{quarks\text{ }\left(  i,j\right)  }\sqrt{2}\left(  \bar{\psi
}_{i}\hat{f}_{ij}\hat{\phi}^{\dagger}L\psi_{j}+\bar{\psi}_{i}\hat{\phi}\hat
{f}_{ji}^{\ast}R\psi_{j}\right)  \label{LYQ}%
\end{equation}
where the summation is over the three different flavors of quark fields for
both $\psi_{i}$ and $\psi_{j}$ and%
\begin{equation}
\hat{f}_{ij}=\left[
\begin{array}
[c]{cc}%
f_{ij}^{u} & 0\\
0 & f_{ij}^{d}%
\end{array}
\right]  . \label{fquark}%
\end{equation}
is a $2\times2$ diagonal matrix. The Yukawa interaction for leptons does not
have terms with mixed generations and is equal to%
\begin{equation}
L_{YL}=-\sum_{leptons\text{ }\left(  i\right)  }\sqrt{2}\left(  \bar{\psi}%
_{i}\hat{f}_{i}\hat{\phi}^{\dagger}L\psi_{i}+\bar{\psi}_{i}\hat{\phi}\hat
{f}_{i}R\psi_{i}\right)  \label{LYL}%
\end{equation}
where the matrix%
\begin{equation}
\hat{f}_{i}=\left[
\begin{array}
[c]{cc}%
f_{i}^{u} & 0\\
0 & f_{i}^{d}%
\end{array}
\right]  \label{flepton}%
\end{equation}
is real and diagonal. Note that from
\[
\hat{\phi}=\frac{1}{\sqrt{2}}\left(  H+v+i\phi_{a}\sigma_{a}\right)  ,
\]
we get%
\[
L_{YL}=\sum_{leptons\text{ }\left(  i\right)  }\bar{\psi}_{i}\left[
\begin{array}
[c]{c}%
-v\hat{f}_{i}-\hat{f}_{i}H\\
+i\left(  \hat{f}_{i}\phi_{a}\sigma_{a}L-\phi_{a}\sigma_{a}\hat{f}%
_{i}R\right)
\end{array}
\right]  \psi_{i}.
\]
The gauge invariant Lagrangian for the fermion fields is
\begin{align}
L_{F}  &  =\sum_{quarks\text{ }\left(  i\right)  }\left[  \bar{\psi}%
_{i}R\left(  i\not D  _{q,L}\right)  L\psi_{i}+\bar{\psi}_{i}L\left(  i\not D
_{q,R}\right)  R\psi_{i}\right]  +L_{YQ}\label{LF}\\
&  +\sum_{leptons\text{ }\left(  i\right)  }\left[  \bar{\psi}_{i}R\left(
i\not D  _{l,L}\right)  L\psi_{i}+\bar{\psi}_{i}L\left(  i\not D
_{l,R}\right)  R\psi_{i}\right]  +L_{YL}\nonumber
\end{align}
which remains gauge invariant even when continued to $n\neq4$. Since
$\gamma^{\mu}L$ is no longer equal to $R\gamma^{\mu}L$ when the polarization
$\mu$ is continued to the extra-4 dimensions, $\bar{\psi}\left(  i\not D
_{L}\right)  L\psi$ and $\bar{\psi}\left(  i\not D  _{L}\right)  R\psi$ depart
from $\bar{\psi}R\left(  i\not D  _{L}\right)  L\psi$ and $\bar{\psi}L\left(
i\not D  _{L}\right)  R\psi$, and are not gauge invariant when $n\neq4$. The
gauge invariant 4 dimensional\ Lagrangian can be conveniently continued to
$n\neq4$ without invalidating gauge symmetry by replacing $\gamma^{\mu}L$ or
$R\gamma^{\mu}$ with $R\gamma^{\mu}L$, and replacing $L\gamma^{\mu}$ or
$\gamma^{\mu}R$ with $L\gamma^{\mu}R$. Let us introduce the notation
$\underline{p^{\mu}}$ for the component of $p^{\mu}$ vector in the first 4
dimensions and the notation $p_{\Delta}^{\mu}$ for the component in the
remaining dimensions. $i.e.,$%
\[
p^{\mu}=\underline{p}^{\mu}+p_{\Delta}^{\mu},
\]
with%
\[
p_{\Delta}^{\mu}=0\text{\ if }\mu\in\left\{  0,1,2,3\right\}
,\underline{\text{ }p}^{\mu}=0\text{\ if }\mu\notin\left\{  0,1,2,3\right\}
.
\]
Likewise, the Dirac matrix $\gamma^{\mu}$ is decomposed as%
\[
\gamma^{\mu}=\underline{\gamma}^{\mu}+\gamma_{\Delta}^{\mu}%
\]
with $\gamma_{\Delta}^{\mu}=0$\ when $\mu\in\left\{  0,1,2,3\right\}  $ and
$\underline{\gamma}^{\mu}=0$ when $\mu\notin\left\{  0,1,2,3\right\}  $.

The free Lagrangian derived from $\left(  \ref{LF}\right)  $ is%
\begin{equation}
L_{F}^{\left(  0\right)  }=\sum_{leptons\text{ }\left(  i\right)  }\bar{\psi
}_{i}\left(  i\underline{\not \partial }-\hat{m}_{i}\right)  \psi_{i}%
+\sum_{quarks\text{ }\left(  i,j\right)  }\bar{\psi}_{i}\left(
i\underline{\not \partial }\delta_{ij}-\left(  \hat{m}_{ij}L+\hat{m}%
_{ji}^{\ast}R\right)  \right)  \psi_{j} \label{LF0}%
\end{equation}
where%
\[
\underline{\not \partial }=\partial_{\mu}\underline{\gamma}^{\mu
}=R\not \partial L+L\not \partial R
\]
and the mass matrices for the fermion fields are%
\begin{equation}
\hat{m}_{i}=\left[
\begin{array}
[c]{cc}%
m_{i}^{u} & 0\\
0 & m_{i}^{d}%
\end{array}
\right]  =v\hat{f}_{i} \label{fml}%
\end{equation}
and%
\begin{equation}
\hat{m}_{ij}=\left[
\begin{array}
[c]{cc}%
m_{ij}^{u} & 0\\
0 & m_{ij}^{d}%
\end{array}
\right]  =v\hat{f}_{ij} \label{fmq}%
\end{equation}
The lepton masses are $m_{i}^{u}=vf_{i}^{u}$ and $m_{i}^{d}=vf_{i}^{d}$ for
$\psi_{i}^{u}$ and $\psi_{i}^{d}$ respectively.

The fermion propagator corresponding to the free Lagrangian $\left(
\ref{LF0}\right)  $ in the momentum space is%
\begin{equation}
\frac{i}{\underline{\not p  }-m} \label{prop4d}%
\end{equation}
which is independent of $p_{\Delta}$, the component of the momentum $p$ in the
extra-4 dimensions and cannot be used for perturbative dimensional
calculation. To remedy this, we add the $CP$ invariant but gauge variant term%
\begin{equation}
E_{0}=\bar{\psi}i\not \partial _{\Delta}\psi=\bar{\psi}_{R}i\not \partial
\psi_{L}+\bar{\psi}_{L}i\not \partial \psi_{R} \label{d00}%
\end{equation}
to the Lagrangian of the theory. The theory so defined will have well-behaved
free fermion propagator%
\[
\frac{i}{\not p  -m}%
\]
and can be used to calculate amplitudes perturbatively under dimensional
regularization scheme. But, we also incur a loss of the gauge symmetry due to
$E_{0}$. Because $E_{0}$ vanishes as $n\rightarrow4$, $E_{0}$ does not have
any tree-level contribution. At one or more loop orders, simple $\frac{1}%
{n-4}$ pole factors or higher pole terms may arise from divergent loop
integrals so that the contribution of $E_{0}$ cannot be neglected and
additional local counter terms are required to restore the gauge symmetry.

A simple and straightforward method for obtaining these finite counter terms
has been proposed \cite{ECSM1,ECSM2} with the help of the rightmost
$\gamma_{5}$ scheme in which the dimension $n$ is analytically continued after
all the $\gamma_{5}$ matrices have been moved to the rightmost position. For
any 1-loop Feynman diagram, the amplitude calculated according to the
rightmost $\gamma_{5}$ scheme can be readily compared with that calculated
directly from the Lagrangian under dimensional regularization with $\gamma
_{5}$ defined in $\left(  \ref{g5def}\right)  $. The difference between these
two amplitudes can be straightforwardly calculated and is equal to the
amplitude due to local counter terms that are required to restore BRST gauge\ symmetry.

\section{1-Loop Electroweak $a_{\mu}^{EW}$}

For simplicity, we choose the Feynman gauge in which $\alpha_{B}=\alpha_{W}=1$
for the gauge fixing terms $\left(  \ref{smgf}\right)  $. Four diagrams are
responsible for the 1-loop electroweak contribution. Those diagrams and their
associated amplitudes are listed in Table \ref{tb1}.%

\begin{table}[ht] \centering
\begin{tabular}
[c]{|l|l|}\hline
Diagram & Contribution to $a_{\mu}^{EW}$\\\hline
$%
\raisebox{-0.4299in}{\includegraphics[
height=0.9262in,
width=1.6143in
]%
{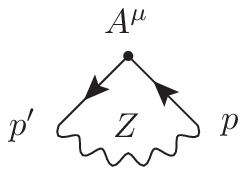}%
}
$ & $\frac{Gm_{\mu}^{2}}{\pi^{2}}\frac{16\left(  \sin^{2}\theta_{W}-\frac
{1}{4}\right)  ^{2}-5}{24\sqrt{2}}$\\\hline
$%
\raisebox{-0.4299in}{\includegraphics[
height=0.9843in,
width=1.6143in
]%
{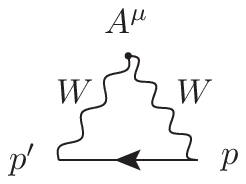}%
}
$ & $\frac{Gm_{\mu}^{2}}{\pi^{2}}\frac{7}{24\sqrt{2}}$\\\hline
$%
\raisebox{-0.4299in}{\includegraphics[
height=0.9843in,
width=1.6143in
]%
{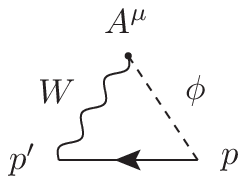}%
}
$ & $0$\\\hline
$%
\raisebox{-0.4299in}{\includegraphics[
height=0.9843in,
width=1.6143in
]%
{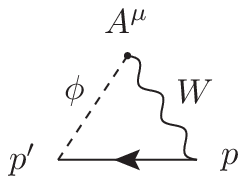}%
}
$ & $\frac{Gm_{\mu}^{2}}{\pi^{2}}\frac{1}{8\sqrt{2}}$\\\hline
\end{tabular}
\caption{1-loop diagrams for $a^{EW}_\mu$}\label{tb1}%
\end{table}%

Adding the elements on second column in Table \ref{tb1} amounts to a total of%
\begin{align}
a_{\mu}^{EW}\left[  \text{1-loop}\right]   &  =\frac{Gm_{\mu}^{2}}{\pi^{2}%
}\frac{16\left(  \sin^{2}\theta_{W}-\frac{1}{4}\right)  ^{2}+5}{24\sqrt{2}%
}\nonumber\\
&  =1.948\times10^{-9} \label{aew1}%
\end{align}
where we have substituted $1.166\times10^{-11}$ $Mev^{-2}$ for the Fermi
coupling constant $G$, $\sin^{2}\theta_{W}=1-\frac{M_{W}^{2}}{M_{Z}^{2}%
}=0.223$, and $105.658$ $MeV$ for the muon mass $m_{u}$ to get the numerical result.

\section{Amplitudes due to Finite Counter Terms at 1-Loop Level}

The finite counter term amplitude is obtained by calculating the difference
arising from moving $\gamma_{5}$ to the rightmost position before continuing
the dimension $n$ \cite{ECSM2}. At 1-loop order, the diagram has to be
divergent by power counting in order to have finite difference between
different ordering of $\gamma_{5}$. Consider the 1-loop fermion self-energy
correction with external muon lines and an internal $W_{1}$ vector meson as
shown below:%
\begin{center}
\includegraphics[
height=0.8782in,
width=2.4999in
]%
{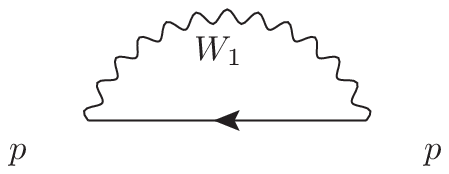}%
\end{center}
The horizontal line signifies an internal fermion line and the wavy line is a
vector meson line. The Feynman amplitude for the self-energy diagram in the
above is%
\[
\Pi_{self\mu}^{W_{1}}=\left(  -i\frac{g_{W}}{2}\right)  ^{2}\int\frac
{d^{n}\ell}{\left(  2\pi\right)  ^{4}}D\left(  W_{\mu},W_{\nu};\ell\right)
R\gamma^{\mu}L\frac{i}{\not \ell+\not p}R\gamma^{\nu}L
\]
where the propagator for $W$ meson is%
\[
D\left(  W_{\mu},W_{\nu};\ell\right)  =-i\frac{g_{\mu\nu}}{\ell^{2}-M_{W}^{2}%
}.
\]
Expand $\frac{1}{\not \ell+\not p}$ as $\frac{1}{\not \ell}-\frac{1}%
{\not \ell}\not p\frac{1}{\not \ell}+...$ and note that the first order term
$\frac{1}{\not \ell}$ does not contribute to the integral $\Gamma
_{self}^{W_{1}}$ from symmetrical integration. We then have (See for Sec. IV
in \cite{ECSM2} details.)
\begin{align*}
\Delta\Pi_{self,\mu}^{W_{1}} &  =i\left(  \frac{g_{W}}{2}\right)  ^{2}%
\Delta\int\frac{d^{n}\ell}{\left(  2\pi\right)  ^{4}}D\left(  W_{\mu},W_{\nu
};\ell\right)  R\gamma^{\mu}L\frac{1}{\not \ell}\not p\frac{1}{\not \ell
}R\gamma^{\nu}L\\
&  =i\left(  \frac{g_{W}}{2}\right)  ^{2}\int\frac{d^{n}\ell}{\left(
2\pi\right)  ^{4}}\frac{D\left(  W_{\mu},W_{\nu};\ell\right)  }{\left(
\ell^{2}\right)  ^{2}}\left(  \gamma^{\mu}\not \ell\not p\not \ell\gamma^{\nu
}L-\underline{\gamma}^{\mu}L\not \ell\not p\not \ell\underline{\gamma}^{\nu
}L\right)  \\
&  =\frac{g_{W}^{2}}{8}\int\frac{d^{n}\ell}{\left(  2\pi\right)  ^{4}}%
\frac{\left(  n-4\right)  }{\left(  \ell^{2}-M_{W}^{2}\right)  ^{2}}%
\not pL=\frac{-ig_{W}^{2}}{64\pi^{2}}R\not pL
\end{align*}

\subsection{Finite Counter Term Amplitude for Self-Energy}

With the external fermion muon and neutrino arranged in a two component matrix
$\left(
\begin{array}
[c]{c}%
\nu_{\mu}\\
\mu
\end{array}
\right)  $ field, the finite counter term contributions with internal $W_{1}$,
$W_{2}$, or $Z$ vector meson can be similarly calculated. The combined total is%

\begin{equation}
\Delta\Pi^{\left(  1\right)  }=-\frac{i}{192\pi^{2}}\left[
\begin{array}
[c]{cc}%
\left(  g_{B}^{2}+3g_{W}^{2}\right)  \not p  L & 0\\
0 & 9\left(  g_{W}^{2}-g_{B}^{2}\right)  \not p  L+16m_{\mu}%
\end{array}
\right]  , \label{Dself}%
\end{equation}
where the first and second diagonal elements correspond to neutrino $\nu_{\mu
}$ and muon $\mu$, respectively. Identify%
\[
\Delta Z_{L}=\frac{1}{64\pi^{2}}\left[
\begin{array}
[c]{cc}%
\left(  g_{B}^{2}+3g_{W}^{2}\right)  & 0\\
0 & 3\left(  g_{W}^{2}-g_{B}^{2}\right)
\end{array}
\right]  ,
\]
and%
\[
\Delta Z_{R}=0.
\]
Then we may write%
\[
\Delta\Pi^{\left(  1\right)  }=-i\left(  \Delta Z_{L}\not p  L+\Delta
Z_{R}\not p  R+\Delta Z_{m}m_{\mu}\right)  .
\]
The fermion propagator then becomes%
\begin{align}
S  &  =\frac{i}{\not p  -m_{\mu}}+\frac{i}{\not p  -m_{\mu}}\Delta\Pi^{\left(
1\right)  }\frac{i}{\not p  -m_{\mu}}\nonumber\\
&  \simeq\left(  \sqrt{Z_{L}}L+\sqrt{Z_{R}}R\right)  \frac{i}{\not p
-\tilde{m}}\left(  \sqrt{Z_{L}}R+\sqrt{Z_{R}}L\right)  \label{fermionS}%
\end{align}
where $Z_{L}=1+\Delta Z_{L}$ and $Z_{R}=1+\Delta Z_{R}$ and $\tilde{m}=\left(
1+\frac{\Delta Z_{L}+\Delta Z_{R}}{2}+\Delta Z_{m}\right)  m_{\mu}$.

\subsection{Finite Counter Term Amplitude for Vertex}

The 1-loop vertex diagram with two external fermion lines is logarithmically
divergent by power counting. As a consequence, amplitude obtained with
rightmost ordering of $\gamma_{5}$ may differ from that obtained with the
$\gamma_{5}$ ordering dictated by the Lagrangian $\left(  \ref{LF}\right)  $
by a finite amount. These finite differences for vertices with external vector
$A$, $Z$, $W$, and scalar $\phi$ are calculated and tabulated in Tables
\ref{tb3}--\ref{tb9}.

\subsubsection{Extended Vertex Factor}

Since an internal fermion line connects one vertex to another, the last factor
$\left(  \sqrt{Z_{L}}R+\sqrt{Z_{R}}L\right)  $ in $\left(  \ref{fermionS}%
\right)  $ can be attributed to the vertex from which the fermion line leaves,
the first factor $\left(  \sqrt{Z_{L}}L+\sqrt{Z_{R}}R\right)  $ in $\left(
\ref{fermionS}\right)  $ can be attributed to the vertex to which the fermion
line flows into, and the fermion propagator stripped off these two factors
effectively becomes a free propagator $\frac{i}{\not p  -\tilde{m}}$. For an
external incoming or outgoing fermion line, $\left(  \sqrt{Z_{L}}R+\sqrt
{Z_{R}}L\right)  $ or $\left(  \sqrt{Z_{L}}L+\sqrt{Z_{R}}R\right)  $ can be
absorbed into the wavefunction renormalization of the external spinor. i.e.,
the fermion propagator can be regarded as $\frac{i}{\not p  -\tilde{m}}$
provided we multiply the vertex factor on the left by $\left(  \sqrt{Z_{L}%
}R+\sqrt{Z_{R}}L\right)  $ and on the right by $\left(  \sqrt{Z_{L}}%
R+\sqrt{Z_{R}}L\right)  $. Diagrammatically, we shall use a large black dot to
denote such an "extended" vertex that includes contributions from all finite
counter terms and the wavefunction normalization factors $\left(  \sqrt{Z_{L}%
}R+\sqrt{Z_{R}}L\right)  $ on the left and $\left(  \sqrt{Z_{L}}L+\sqrt{Z_{R}%
}R\right)  $ on the right. The 1-loop vertex factors for all possible black
dots are calculated and listed in Tables \ref{tb2-1} and \ref{tb2-2}.%

\begin{figure}[tbh]%
\centering
\includegraphics[
height=1.4077in,
width=5.5994in
]%
{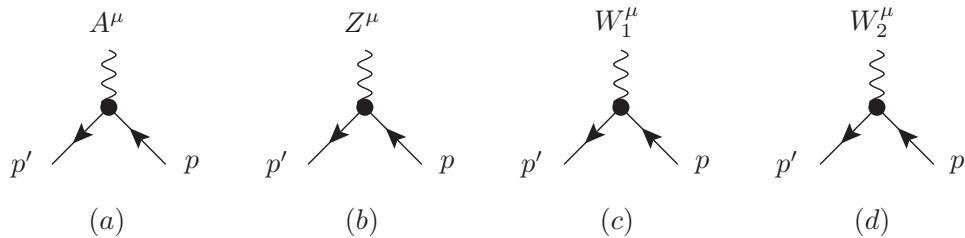}%
\caption{Extended Vertex Diagrams for fermion-vector-fermion}%
\label{fig-evtx-vector}%
\end{figure}

\begin{table}[ht] \centering
\begin{tabular}
[c]{|l|l|}\hline
Figure & $\Gamma^{\mu}-\Gamma_{tree}^{\mu}$\\\hline
\ref{fig-evtx-vector}$(a)$ & $-ie\frac{g_{W}^{2}}{32\pi^{2}}\left[
\begin{array}
[c]{cc}%
1 & 0\\
0 & -1
\end{array}
\right]  \gamma^{\mu}L$\\\hline
\ref{fig-evtx-vector}$(b)$ & $\frac{i\sqrt{g_{W}^{2}+g_{B}^{2}}}{128\pi^{2}%
}\left(
\begin{array}
[c]{c}%
\frac{3g_{W}^{4}+g_{B}^{4}}{g_{W}^{2}+g_{B}^{2}}\left[
\begin{array}
[c]{cc}%
1 & 0\\
0 & -1
\end{array}
\right]  \gamma^{\mu}L\\
+g_{B}^{2}\left[
\begin{array}
[c]{cc}%
0 & 0\\
0 & 8\gamma^{\mu}-12\gamma^{\mu}L
\end{array}
\right]
\end{array}
\right)  $\\\hline
\ref{fig-evtx-vector}$(c)$ & $\frac{3ig_{W}\left(  g_{W}^{2}+g_{B}^{2}\right)
}{128\pi^{2}}\left[
\begin{array}
[c]{cc}%
0 & 1\\
1 & 0
\end{array}
\right]  \gamma^{\mu}L$\\\hline
\ref{fig-evtx-vector}$(d)$ & $\frac{3ig_{W}\left(  g_{W}^{2}+g_{B}^{2}\right)
}{128\pi^{2}}\left[
\begin{array}
[c]{cc}%
0 & -i\\
i & 0
\end{array}
\right]  \gamma^{\mu}L$\\\hline
\end{tabular}
\caption{Extended Vertex Factors for fermion-vector-fermion at 1-loop order}\label{tb2-1}%
\end{table}%

\begin{figure}[tbh]%
\centering
\includegraphics[
height=1.4077in,
width=4.2775in
]%
{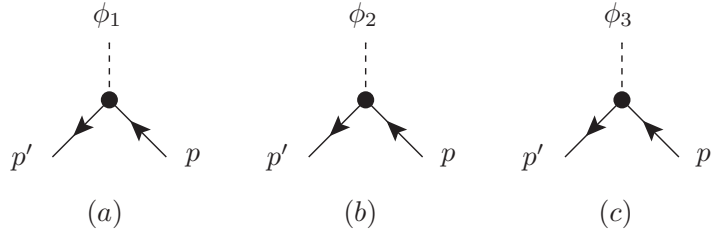}%
\caption{Extended Vertex Diagrams for fermion-scalar-fermion}%
\label{fig-evtx-fi}%
\end{figure}

\begin{table}[ht] \centering
\begin{tabular}
[c]{|l|l|}\hline
Figure & $\Gamma-\Gamma_{tree}$\\\hline
\ref{fig-evtx-fi}$\left(  a\right)  $ & $f\frac{\left(  g_{W}^{2}-14g_{B}%
^{2}\right)  }{128\pi^{2}}\left[
\begin{array}
[c]{cc}%
0 & R\\
-L & 0
\end{array}
\right]  $\\\hline
\ref{fig-evtx-fi}$\left(  b\right)  $ & $-if\frac{\left(  g_{W}^{2}%
-14g_{B}^{2}\right)  }{128\pi^{2}}\left[
\begin{array}
[c]{cc}%
0 & R\\
L & 0
\end{array}
\right]  $\\\hline
\ref{fig-evtx-fi}$\left(  c\right)  $ & $f\frac{\left(  g_{W}^{2}-43g_{B}%
^{2}\right)  }{128\pi^{2}}\left[
\begin{array}
[c]{cc}%
0 & 0\\
0 & \left(  L-R\right)
\end{array}
\right]  $\\\hline
\end{tabular}
\caption{Extended Vertex Factors for fermion-scalar-fermion}\label{tb2-2}%
\end{table}%

\section{Finite-counter-term Contribution to $a_{\mu}^{EW}$}

To obtain the finite-counter-term contribution for $a_{\mu}^{EW}$, replace the
fermion vertex factors in each of the four diagram of Table \ref{tb1} with the
large black dot representing the extended vertex as shown in Table \ref{tb2}.
The amplitude with $a_{\mu}^{EW}\left[  \text{tree}\right]  $ and $a_{\mu
}^{EW}\left[  \text{1-loop}\right]  $ subtracted out is the
finite-counter-term to $a_{\mu}^{EW}$.%

\begin{table}[ht] \centering
\begin{tabular}
[c]{|l|l|}\hline
Diagram & Finite-counter-term Contribution to $a_{\mu}^{EW}$\\\hline%
\raisebox{-0.4289in}{\includegraphics[
height=0.9262in,
width=1.6143in
]%
{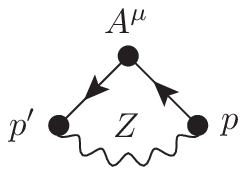}%
}
& $\left.
\begin{array}
[c]{c}%
\frac{\alpha Gm_{\mu}^{2}}{12\sqrt{2}\pi^{3}}\csc^{2}\left(  2\theta
_{W}\right)  \\
\times\left(  6\sin^{6}\left(  \theta_{W}\right)  -7\sin^{4}\left(  \theta
_{W}\right)  +11\sin^{2}\left(  \theta_{W}^{2}\right)  +4\right)
\end{array}
\right.  $\\\hline
$%
\raisebox{-0.4289in}{\includegraphics[
height=0.9843in,
width=1.6143in
]%
{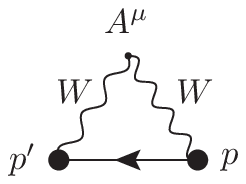}%
}
$ & $-\frac{7\alpha Gm_{\mu}^{2}}{17\sqrt{2}\pi^{3}}\csc^{2}\left(
2\theta_{W}\right)  $\\\hline
$%
\raisebox{-0.4289in}{\includegraphics[
height=0.9843in,
width=1.6143in
]%
{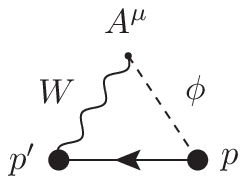}%
}
$ & $\frac{5\alpha Gm_{\mu}^{2}}{128\sqrt{2}\pi^{3}}\csc^{2}\left(
2\theta_{W}\right)  \left(  3\cos\left(  2\theta_{W}\right)  -5\right)
$\\\hline
$%
\raisebox{-0.4289in}{\includegraphics[
height=0.9843in,
width=1.6143in
]%
{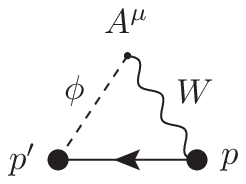}%
}
$ & $0$\\\hline
\end{tabular}
\caption{Finite-counter-term Diagrams}\label{tb2}%
\end{table}%

$a_{\mu}^{EW}\left[  \text{Finite Counter term}\right]  $ is obtained by
summing over the second column. The result is
\[
-\frac{\alpha Gm_{\mu}^{2}}{384\sqrt{2}\pi^{3}}\csc^{2}\left(  2\theta
_{W}\right)  \left(  6\cos\left(  6\theta_{W}\right)  -8\cos\left(
4\theta_{W}\right)  +109\cos\left(  2\theta_{W}\right)  -37\right)
\]
which is numerically evaluated to
\begin{equation}
a_{\mu}^{EW}\left[  \text{Finite Counter term}\right]  =-1.67541\times10^{-12}
\label{afct}%
\end{equation}
where we have set $\alpha=\frac{1}{137.036}$.

\section{Conclusion}

In the dimensional regularization scheme, simply removing the pole terms from
the amplitudes of 1-loop diagrams does not yield renormalized amplitudes that
satisfy the BRST gauge symmetry. Instead, some finite renormalization terms
have to be added. The renormalized amplitudes for all 1-loop diagrams
calculated in the straightforward dimensional regularization scheme with
finite counter term renormalization are equal to those obtained in the
rightmost $\gamma_{5}$ scheme. This means we can be spared the tedious finite
renormalization procedures if the rightmost $\gamma_{5}$ scheme is adopted as
we have shown in this paper for the evaluation of $a_{\mu}=\frac{g_{\mu}-2}%
{2}$, where $g_{\mu}$ is the muon gyromagnetic ratio.

The $a_{\mu}^{EW}\left[  \text{Finite Counter term}\right]  =-1.67541\times
10^{-12}$ we have arrived at in $\left(  \ref{afct}\right)  $ due to the
finite counter terms arising from electroweak interaction in the standard
model is only about one-thousandth of the $a_{\mu}^{EW}\left[  \text{1-loop}%
\right]  =1.948\times10^{-9}$ in $\left(  \ref{aew1}\right)  $ or the
difference $\Delta a_{\mu}=2.81\times10^{-9}$ between experiment and theory in
$\left(  \ref{gdiff}\right)  $. The finite counter term contribution to the
muon magnetic moment is therefore not significant enough to account for the
discrepancy between experiment and theoretical prediction by standard model.%

\appendix\appendixpage

\section{Counter term Amplitudes due to Vertex Diagrams}

Vertex diagrams that are relevant to the calculation of the finite counter
terms contributing to the muon anomalous magnetic moment are drawn below in
Figures \ref{fig-vtxA}--\ref{fig-vtxf3} with corresponding amplitudes listed
in Tables \ref{tb3}--\ref{tb9}.%

\begin{figure}[tbh]%
\centering
\includegraphics[
height=1.261in,
width=4.2775in
]%
{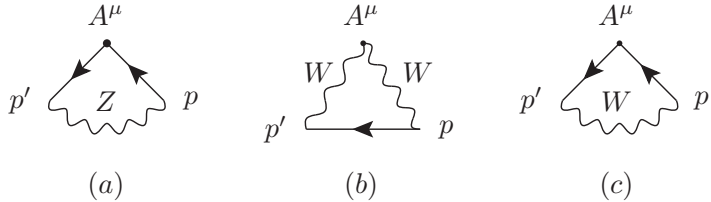}%
\caption{ Vertex Diagrams for $\bar{\Psi}A^{\mu}\Psi$}%
\label{fig-vtxA}%
\end{figure}

\begin{table}[ht] \centering
\begin{tabular}
[c]{|l|l|}\hline
Figure & $\Delta\Gamma^{A^{\mu}}$\\\hline
\ref{fig-vtxA}$(a)$ & $\frac{ie\left(  g_{W}^{2}-3g_{B}^{2}\right)  }%
{64\pi^{2}}\gamma^{\mu}\left[
\begin{array}
[c]{cc}%
0 & 0\\
0 & 1
\end{array}
\right]  L$\\\hline
\ref{fig-vtxA}$(b)$ & $\frac{-ieg_{W}^{2}}{16\pi^{2}}\gamma^{\mu}\left[
\begin{array}
[c]{cc}%
1 & 0\\
0 & -1
\end{array}
\right]  L$\\\hline
\ref{fig-vtxA}$(c)$ & $\frac{ieg_{W}^{2}}{32\pi^{2}}\gamma^{\mu}\left[
\begin{array}
[c]{cc}%
1 & 0\\
0 & 0
\end{array}
\right]  L$\\\hline
\end{tabular}
\caption{Counter term Amplitude for fermion-$A^\mu$-fermion vertex}\label{tb3}%
\end{table}%

\clearpage

\begin{figure}[tbh]%
\centering
\includegraphics[
height=1.4077in,
width=5.5994in
]%
{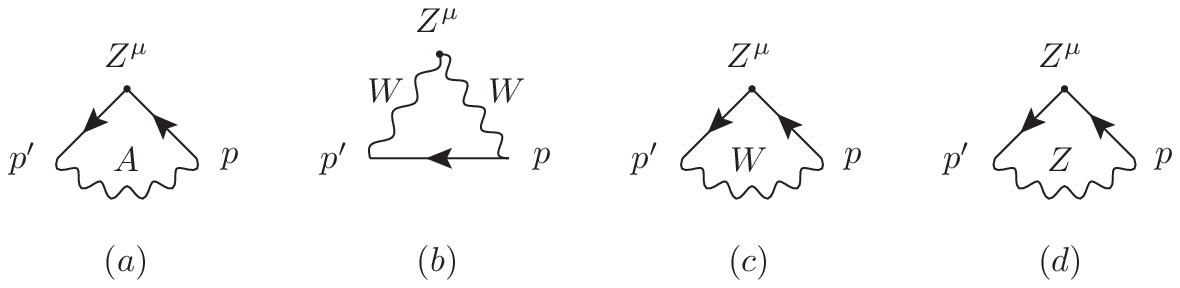}%
\caption{ Vertex Diagrams for $\bar{\Psi}Z^{\mu}\Psi$}%
\label{fig-vtxZ}%
\end{figure}

\begin{table}[ht] \centering
\begin{tabular}
[c]{|l|l|}\hline
Figure & $\Delta\Gamma^{Z^{\mu}}$\\\hline
\ref{fig-vtxZ}$(a)$ & $\frac{ie^{2}\sqrt{g_{B}^{2}+g_{W}^{2}}}{16\pi^{2}%
}\gamma^{\mu}\left[
\begin{array}
[c]{cc}%
0 & 0\\
0 & R-L
\end{array}
\right]  $\\\hline
\ref{fig-vtxZ}$(b)$ & $\frac{ieg_{W}^{3}}{16\pi^{2}}\gamma^{\mu}\left[
\begin{array}
[c]{cc}%
1 & 0\\
0 & -1
\end{array}
\right]  L$\\\hline
\ref{fig-vtxZ}$(c)$ & $\frac{ig_{W}^{2}}{32\pi^{2}\sqrt{g_{B}^{2}+g_{W}^{2}}%
}\gamma^{\mu}\left[
\begin{array}
[c]{cc}%
-g_{W}^{2} & 0\\
0 & \left(  g_{W}^{2}+g_{B}^{2}\right)
\end{array}
\right]  L$\\\hline
\ref{fig-vtxZ}$(d)$ & $\frac{i\left(  g_{B}^{2}+g_{W}^{2}\right)  ^{\frac
{3}{2}}}{64\pi^{2}}\gamma^{\mu}\left[
\begin{array}
[c]{cc}%
L & 0\\
0 & \frac{4g_{B}^{4}-\left(  g_{W}^{4}-3g_{B}^{2}g_{W}^{2}+8g_{B}^{4}\right)
L}{\left(  g_{B}^{2}+g_{W}^{2}\right)  ^{2}}%
\end{array}
\right]  $\\\hline
\end{tabular}
\caption{Counter term Amplitude for fermion-$Z^\mu$-fermion vertex}\label{tb4}%
\end{table}%

\clearpage

\begin{figure}[tbh]%
\centering
\includegraphics[
height=1.2684in,
width=6.7229in
]%
{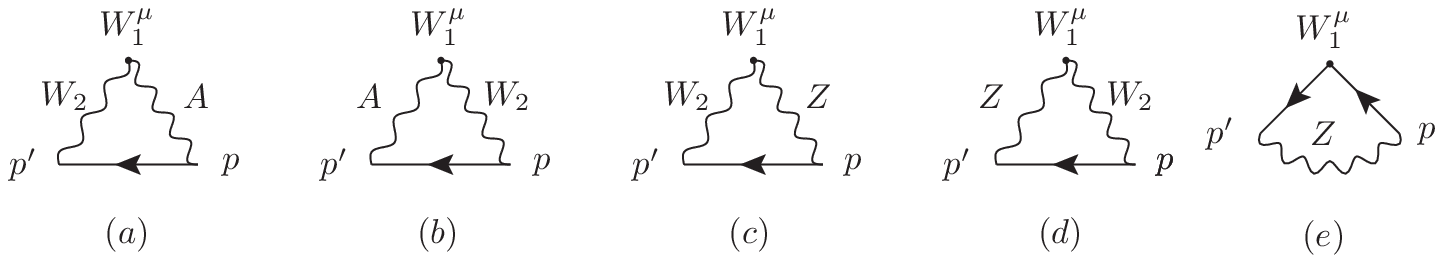}%
\caption{ Vertex Diagrams for $\bar{\Psi}W_{1}^{\mu}\Psi$}%
\label{fig-vtxw1}%
\end{figure}
\begin{table}[ht] \centering
\begin{tabular}
[c]{|l|l|}\hline
Figure & $\Delta\Gamma^{W_{1}^{\mu}}$\\\hline
\ref{fig-vtxw1}$(a)$ & $\frac{3ie^{2}g_{W}}{32\pi^{2}}\gamma^{\mu}\left[
\begin{array}
[c]{cc}%
0 & 1\\
0 & 0
\end{array}
\right]  L$\\\hline
\ref{fig-vtxw1}$(b)$ & $\frac{3ie^{2}g_{W}}{32\pi^{2}}\gamma^{\mu}\left[
\begin{array}
[c]{cc}%
0 & 0\\
1 & 0
\end{array}
\right]  L$\\\hline
\ref{fig-vtxw1}$(c)$ & $\frac{ie^{2}g_{W}}{64\pi^{2}g_{B}}\gamma^{\mu}\left[
\begin{array}
[c]{cc}%
0 & 2g_{W}^{2}-g_{B}^{2}\\
2\left(  g_{W}^{2}+g_{B}^{2}\right)   & 0
\end{array}
\right]  L$\\\hline
\ref{fig-vtxw1}$(d)$ & $\frac{ie^{2}g_{W}}{64\pi^{2}g_{B}}\gamma^{\mu}\left[
\begin{array}
[c]{cc}%
0 & 2\left(  g_{W}^{2}+g_{B}^{2}\right)  \\
2g_{W}^{2}-g_{B}^{2} & 0
\end{array}
\right]  $\\\hline
\ref{fig-vtxw1}$(e)$ & $\frac{-ig_{W}\left(  g_{W}^{2}-g_{B}^{2}\right)
}{64\pi^{2}}\gamma^{\mu}\left[
\begin{array}
[c]{cc}%
0 & 1\\
1 & 0
\end{array}
\right]  L$\\\hline
\end{tabular}
\caption{Counter term Amplitude for fermion-$W_1^\mu$-fermion vertex}\label{tb5}%
\end{table}%

\clearpage

\begin{figure}[tbh]%
\centering
\includegraphics[
height=1.2601in,
width=6.6399in
]%
{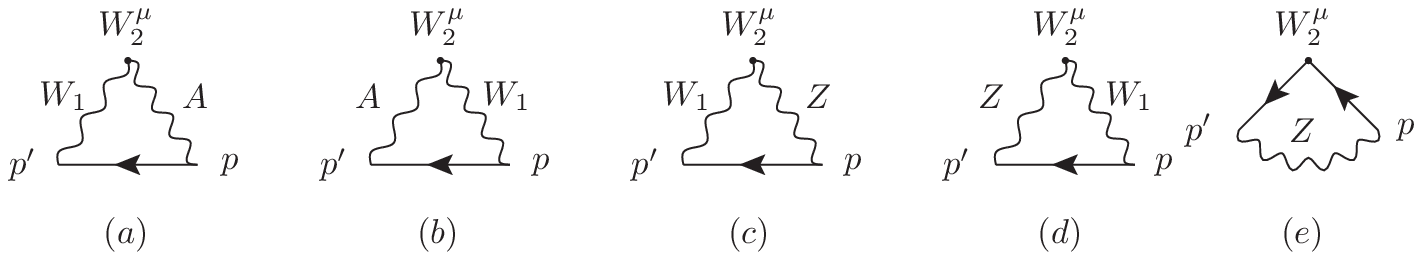}%
\caption{ Vertex Diagrams for $\bar{\Psi}W_{2}^{\mu}\Psi$}%
\label{fig-vtxw2}%
\end{figure}

\begin{table}[ht] \centering
\begin{tabular}
[c]{|l|l|}\hline
Figure & $\Delta\Gamma^{W_{2}^{\mu}}$\\\hline
\ref{fig-vtxw2}$(a)$ & $\frac{3ie^{2}g_{W}}{32\pi^{2}}\gamma^{\mu}\left[
\begin{array}
[c]{cc}%
0 & -i\\
0 & 0
\end{array}
\right]  L$\\\hline
\ref{fig-vtxw2}$(b)$ & $\frac{3ie^{2}g_{W}}{32\pi^{2}}\gamma^{\mu}\left[
\begin{array}
[c]{cc}%
0 & 0\\
i & 0
\end{array}
\right]  L$\\\hline
\ref{fig-vtxw2}$(c)$ & $\frac{e^{2}g_{W}}{64\pi^{2}g_{B}}\gamma^{\mu}\left[
\begin{array}
[c]{cc}%
0 & 2g_{W}^{2}-g_{B}^{2}\\
-2\left(  g_{W}^{2}+g_{B}^{2}\right)  & 0
\end{array}
\right]  L$\\\hline
\ref{fig-vtxw2}$(d)$ & $\frac{e^{2}g_{W}}{64\pi^{2}g_{B}}\gamma^{\mu}\left[
\begin{array}
[c]{cc}%
0 & 2\left(  g_{W}^{2}+g_{B}^{2}\right) \\
-2g_{W}^{2}+g_{B}^{2} & 0
\end{array}
\right]  L$\\\hline
\ref{fig-vtxw2}$(e)$ & $\frac{-ig_{W}\left(  g_{W}^{2}-g_{B}^{2}\right)
}{64\pi^{2}}\gamma^{\mu}\left[
\begin{array}
[c]{cc}%
0 & -i\\
i & 0
\end{array}
\right]  L$\\\hline
\end{tabular}
\caption{Counter term Amplitude for fermion-$W_2^\mu$-fermion vertex}\label{tb6}%
\end{table}%

\clearpage

\begin{figure}[tbh]%
\centering
\includegraphics[
height=1.261in,
width=5.746in
]%
{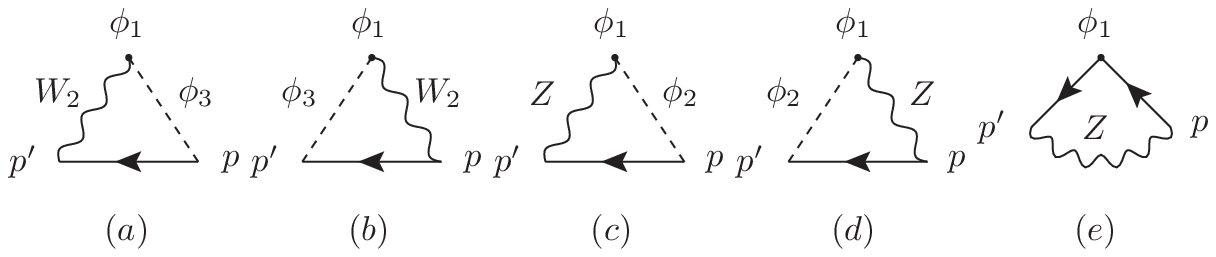}%
\caption{ Vertex Diagrams for $\bar{\Psi}\phi_{1}\Psi$}%
\label{fig-vtxf1}%
\end{figure}

\begin{table}[ht] \centering
\begin{tabular}
[c]{|l|l|}\hline
Figure & $\Delta\Gamma^{\phi_{1}}$\\\hline
\ref{fig-vtxf1}$(a)$ & $\frac{fg_{W}^{2}}{128\pi^{2}}\left[
\begin{array}
[c]{cc}%
0 & -R\\
0 & 0
\end{array}
\right]  $\\\hline
\ref{fig-vtxf1}$(b)$ & $\frac{fg_{W}^{2}}{128\pi^{2}}\left[
\begin{array}
[c]{cc}%
0 & 0\\
L & 0
\end{array}
\right]  $\\\hline
\ref{fig-vtxf1}$(c)$ & $\frac{f\left(  g_{W}^{2}-g_{B}^{2}\right)  }%
{128\pi^{2}}\left[
\begin{array}
[c]{cc}%
0 & -R\\
0 & 0
\end{array}
\right]  $\\\hline
\ref{fig-vtxf1}$(d)$ & $\frac{f\left(  g_{W}^{2}-g_{B}^{2}\right)  }%
{128\pi^{2}}\left[
\begin{array}
[c]{cc}%
0 & 0\\
L & 0
\end{array}
\right]  $\\\hline
\ref{fig-vtxf1}$(e)$ & $\frac{fg_{B}^{2}}{8\pi^{2}}\left[
\begin{array}
[c]{cc}%
0 & -R\\
L & 0
\end{array}
\right]  $\\\hline
\end{tabular}
\caption{Counter term Amplitude for fermion-$\phi_1$-fermion vertex}\label{tb7}%
\end{table}%

\clearpage

\begin{figure}[tbh]%
\centering
\includegraphics[
height=1.261in,
width=5.746in
]%
{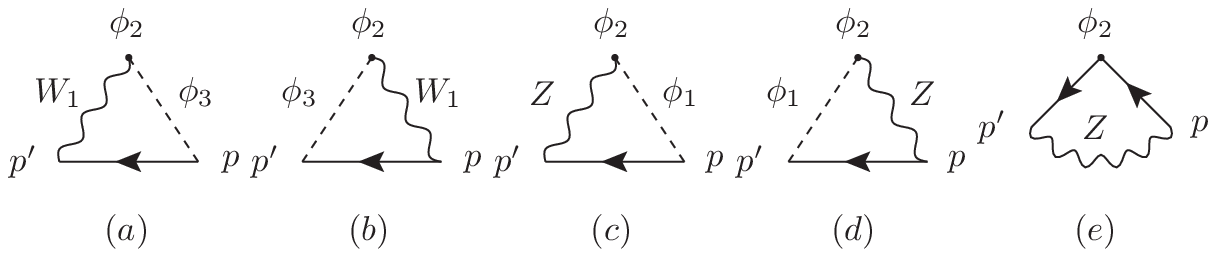}%
\caption{ Vertex Diagrams for $\bar{\Psi}\phi_{2}\Psi$}%
\label{fig-vtxf2}%
\end{figure}

\begin{table}[ht] \centering
\begin{tabular}
[c]{|l|l|}\hline
Figure & $\Delta\Gamma^{\phi_{2}}$\\\hline
\ref{fig-vtxf2}$(a)$ & $\frac{ifg_{W}^{2}}{128\pi^{2}}\left[
\begin{array}
[c]{cc}%
0 & R\\
0 & 0
\end{array}
\right]  $\\\hline
\ref{fig-vtxf2}$(b)$ & $\frac{ifg_{W}^{2}}{128\pi^{2}}\left[
\begin{array}
[c]{cc}%
0 & 0\\
L & 0
\end{array}
\right]  $\\\hline
\ref{fig-vtxf2}$(c)$ & $\frac{if\left(  g_{W}^{2}-g_{B}^{2}\right)  }%
{128\pi^{2}}\left[
\begin{array}
[c]{cc}%
0 & R\\
0 & 0
\end{array}
\right]  $\\\hline
\ref{fig-vtxf2}$(d)$ & $\frac{if\left(  g_{W}^{2}-g_{B}^{2}\right)  }%
{128\pi^{2}}\left[
\begin{array}
[c]{cc}%
0 & 0\\
L & 0
\end{array}
\right]  $\\\hline
\ref{fig-vtxf2}$(e)$ & $\frac{ifg_{B}^{2}}{8\pi^{2}}\left[
\begin{array}
[c]{cc}%
0 & R\\
L & 0
\end{array}
\right]  $\\\hline
\end{tabular}
\caption{Counter term Amplitude for fermion-$\phi_2$-fermion vertex}\label{tb8}%
\end{table}%

\clearpage

\begin{figure}[tbh]%
\centering
\includegraphics[
height=2.5977in,
width=4.8559in
]%
{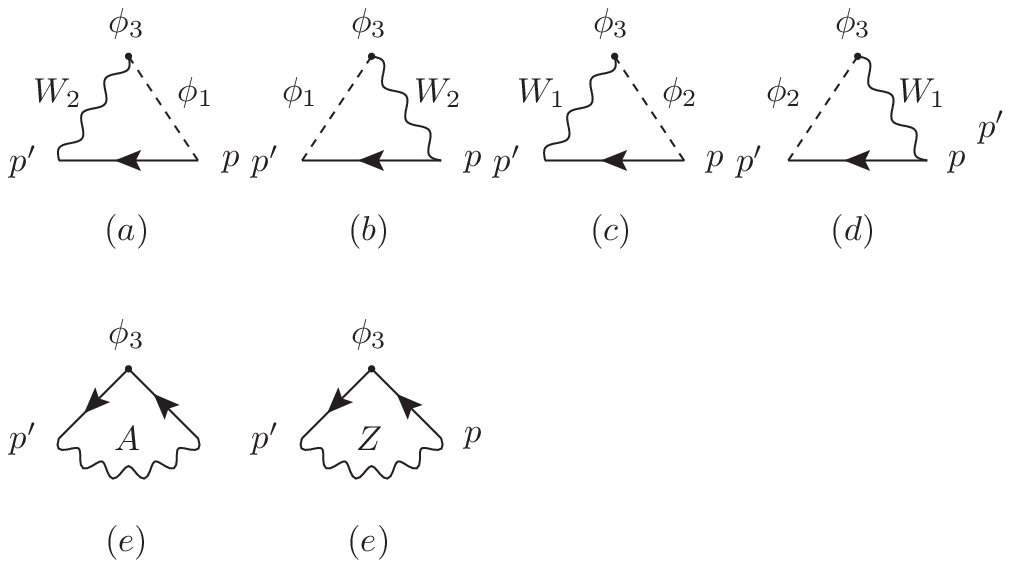}%
\caption{ Vertex Diagrams for $\bar{\Psi}\phi_{3}\Psi$}%
\label{fig-vtxf3}%
\end{figure}

\begin{table}[ht] \centering
\begin{tabular}
[c]{|l|l|}\hline
Figure & $\Delta\Gamma^{\phi_{3}}$\\\hline
\ref{fig-vtxf3}$(a)$ & $\frac{fg_{W}^{2}}{128\pi^{2}}\left[
\begin{array}
[c]{cc}%
0 & 0\\
0 & R
\end{array}
\right]  $\\\hline
\ref{fig-vtxf3}$(b)$ & $\frac{fg_{W}^{2}}{128\pi^{2}}\left[
\begin{array}
[c]{cc}%
0 & 0\\
0 & -L
\end{array}
\right]  $\\\hline
\ref{fig-vtxf3}$(c)$ & $\frac{fg_{W}^{2}}{128\pi^{2}}\left[
\begin{array}
[c]{cc}%
0 & 0\\
0 & R
\end{array}
\right]  $\\\hline
\ref{fig-vtxf3}$(d)$ & $\frac{fg_{W}^{2}}{128\pi^{2}}\left[
\begin{array}
[c]{cc}%
0 & 0\\
0 & -L
\end{array}
\right]  $\\\hline
\ref{fig-vtxf3}$(e)$ & $\frac{e^{2}f}{2\pi^{2}}\left[
\begin{array}
[c]{cc}%
0 & 0\\
0 & R-L
\end{array}
\right]  $\\\hline
\ref{fig-vtxf3}$(f)$ & $\frac{e^{2}f}{16\pi^{2}}\left(  3-5\frac{g_{B}^{2}%
}{g_{W}^{2}}\right)  \left[
\begin{array}
[c]{cc}%
0 & 0\\
0 & R-L
\end{array}
\right]  $\\\hline
\end{tabular}
\caption{Counter term Amplitude for fermion-$\phi_3$-fermion vertex}\label{tb9}%
\end{table}%

\end{document}